# Evolution from unconventional spin density wave to superconductivity and a novel gap-like phase in NaFe$_{1-x}$Co$_x$As

Xiaodong Zhou[1,*], Peng Cai[1,*], Aifeng Wang[2], Wei Ruan[1], Cun Ye[1], Xianhui Chen[2], Yizhuang You[3], Zheng-Yu Weng[3], and Yayu Wang[1,†]

[1]*State Key Laboratory of Low Dimensional Quantum Physics, Department of Physics, Tsinghua University, Beijing 100084, P. R. China*

[2]*Hefei National Laboratory for Physical Science at Microscale and Department of Physics, University of Science and Technology of China, Hefei, Anhui 230026, P.R. China*

[3]*Institute for Advanced Study, Tsinghua University, Beijing 100084, P. R. China*

\* *These authors contributed equally to this work.*

[†] Email: yayuwang@tsinghua.edu.cn

**Similar to the cuprate high $T_C$ superconductors, the iron pnictide superconductors also lie in close proximity to a magnetically ordered phase[1,2]. A central debate concerning the superconducting mechanism is whether the local magnetic moments play an indispensable role or the itinerant electron description is sufficient[3-8]. A key step for resolving this issue is to acquire a comprehensive picture regarding the nature of various phases and interactions in the iron compounds. Here we report the doping, temperature, and spatial evolutions of the electronic structure of $NaFe_{1-x}Co_xAs$ studied by scanning tunneling microscopy. The spin density wave gap in the parent state is observed for the first time, which shows a strongly asymmetric lineshape that is incompatible with the conventional Fermi surface nesting scenario. The optimally doped sample exhibits a single, symmetric energy gap, but in the overdoped regime another asymmetric gap-like feature emerges near the Fermi level. This novel gap-like phase coexists with superconductivity in the ground state, persists deep into the normal state, and shows strong spatial variations. The characteristics of the three distinct low energy states, in conjunction with the peculiar high energy spectra, suggest that the coupling between the local moments and itinerant electrons is the fundamental driving force for the phases and phase transitions in the iron pnictides.**

Currently there are drastically different views regarding the physics of iron-based superconductors. The existence of metallic conduction and well-defined Fermi surface (FS) in the parent state prompts the viewpoint that the magnetic ordering is a spin density wave (SDW) induced by FS nesting of itinerant electrons, and upon doping the pair scattering between the electron- and hole-like FS pockets may lead to superconductivity[3,6]. On the contrary, another group of thought ascribes the magnetism to antiferromagnetic (AF) ordering of local moments, which gives better description of the spin wave spectrum, and the magnetic interaction may directly be responsible for Cooper pairing in the doped case[4,5]. To reconcile the seemingly conflicting facts, another theoretical model proposes that the local moments and itinerant electrons may coexist and the coupling between them gives rise to both the SDW and superconducting (SC) phases[7,8].

A promising strategy for differentiating these theoretical models is to clarify the nature of each microscopic phase in the iron pnictides and establish a unified underlying picture. Owing to its ability to probe the atomic scale electronic structure for both the occupied and empty states, scanning tunneling microscopy (STM) has played a crucial role in fulfilling this task[9-12]. However, up to date a comprehensive picture about the phase diagram of the iron pnictide superconductors is still lacking. For example, the SDW gap in the parent state has never been directly observed, and its spectroscopic features remain undisclosed. There are also controversies regarding the coexisting or competing orders in the SC state, and whether there exists a pseudogap phase[13,14]. Furthermore, the electronic signatures of the existence of local magnetic moment and its coupling with the itinerant electrons are still elusive.

In this work we report STM investigations of the electronic structures of the 111-type $NaFe_{1-x}Co_xAs$ system, which has a mirror-symmetric layered structure as drawn in Fig. 1a. The crystal is expected to cleave between two weakly bonded Na intercalation layers, leading to a Na-terminated square lattice as shown in Fig. 1b. Fig. 1c displays the surface topography of a cleaved parent NaFeAs crystal. The surface is clean and flat over a large area, but there are nanometer-sized dark pits with unknown origin (Fig. S1). The inset shows a high resolution zoom-in image on a defect-free area, which reveals a square lattice with lattice constant $a \sim 4$ Å. This is consistent with the expected surface structure shown in Fig. 1b and previous STM images on similar 111 pnictides[15,16]. The absence of surface reconstruction and extrinsic surface states suggest that the STM results obtained here can reflect the intrinsic bulk electronic structure of $NaFe_{1-x}Co_xAs$.

Figure 1d illustrates the electronic phase diagram of $NaFe_{1-x}Co_xAs$. With increasing Co content, hence the density of doped electrons, the system evolves from the SDW or AF ordered parent state to a dome-shaped SC regime and then to the non-SC metallic phase. Fig. 1e shows the resistivity ($\rho$) vs. temperature ($T$) curves of the five $NaFe_{1-x}Co_xAs$ crystals studied here, which cover each representative regime as marked by the solid symbols in the phase diagram. In the parent NaFeAs although $\rho$ reaches zero for $T < 10$K, it has been shown that the superconductivity is filamentary rather than a bulk phenomenon[17]. Therefore its $T_C$ is allocated to zero in the phase diagram.

Figure 2 displays the spatially-averaged $dI/dV$ (differential conductance) spectroscopy, which is approximately proportional to the electron density of state (DOS). The upper panels show the $dI/dV$ curves taken at the base temperature $T = 5$ K, and the lower panels are that taken at varied temperatures. In the ground state of parent NaFeAs ($x = 0$), we observe a well-defined energy gap formed around $E_F$ (Fig. 2a). With increasing $T$ this gap is gradually filled up and closes at $T = 40$ K (Fig. 2b), which is the SDW transition temperature $T_{SDW}$ determined by various experimental probes[17,18]. This energy gap is thus the highly anticipated SDW gap, which has never been directly observed before by STM or angle-resolved photoemission spectroscopy (ARPES) in parent pnictides, in spite of implications from infrared spectroscopy[19]. The low $T$ spectrum shown in Fig. 2a reveals three important features of the SDW gap. Firstly, the gap amplitude defined from the distance between the two peaks is $2\Delta_{SDW} = 33$ meV, which gives an anomalously large $2\Delta_{SDW}/k_BT_{SDW}$ ratio of 9.5. Secondly, the gap structure is highly asymmetric with respect to $E_F$, apparently breaking the particle-hole symmetry. The spectrum is tilted towards positive bias with the gap bottom located at 4 meV above $E_F$ and the two peaks located at -15 meV and +18 meV respectively. Thirdly, there is a large residual DOS even for $T$ well below $T_{SDW}$, indicating that the FS is only partially gapped by the SDW order.

Figure 2c displays the *dI/dV* curve taken on the optimally doped $x = 0.028$ sample, which exhibits a sharp SC gap with well-defined coherence peaks. Unlike the SDW gap shown in Fig. 2a, the SC gap here is highly symmetric with respect to $E_F$. The gap amplitude defined from the distance between the two coherence peaks is $2\Delta_{SC} = 11$ meV, leading to a ratio $2\Delta_{SC}/k_BT_C = 6.4$. This value is significantly larger than that for weak coupling BCS superconductors, but is close to previous STM studies on other iron pnictides[20]. Fig. 2d shows that the SC gap closes right at $T_C = 20$ K, and in the normal state the *dI/dV* curves only present weak spectroscopic features.

The electronic structure of the overdoped $x = 0.061$ sample displays a more complex pattern. At $T = 5$ K the energy gap has a minimum right at $E_F$, but the peak on the negative bias side has a much larger amplitude and lies further away from $E_F$ than that on the positive side (Fig. 2e). With increasing $T$ the sharp gap feature near $E_F$ becomes shallower and disappears above $T_C = 13$ K (Fig. 2f), confirming that it is the SC gap. However, another asymmetric gap-like feature with a peak at negative bias and a dip near $E_F$ persists to the normal state and only disappears gradually at much higher *T*. The $x = 0.075$ sample shows a similar behavior, except that the SC gap feature at $T = 5$ K (Fig. 2g) becomes even weaker. The SC gap in these two samples can be extracted by dividing the *dI/dV* curve taken at $T = 5$ K by that obtained just above $T_C$ (Fig. S2). The resulting spectra are very similar to that on the optimally doped $x = 0.028$ sample, with a highly symmetric SC gap and systematically decreasing gap size with increasing *x*. Similarly, the novel gap-like feature can be further enhanced when the high *T* background taken at 70 K is divided out (Fig. S3). The *T* evolution of the novel-gap like feature is highly analogous to the pseudogap in underdoped cuprates[21]. It persists deep into the normal state and vanishes smoothly near $T = 55$ K.

We gain more insight into the novel gap-like phase by taking advantage of the spatial resolution of STM. Fig. 3a shows the topography of the $x = 0.061$ sample, which becomes more disordered than the parent NaFeAs due to Co doping (Fig. S4). Fig. 3b displays the *dI/dV* spectra taken at $T = 5$ K along the line drawn in Fig. 3a. The *dI/dV* curves of the majority phase (the bright area in the topography) look similar to the one shown in Fig. 2e. In the minority dark area, however, the energy scales of the SC gap and large asymmetric gap become well-separated. The coexistence of the two distinct phases becomes more apparent. Fig. 3c displays the *dI/dV* spectra taken along the same line at $T = 16$K, where the SC gap closes but the large asymmetric gap remains and shows strong spatial variations. Fig. 3d shows the normalized *dI/dV* spectra, which clearly demonstrates that the SC gap is particle-hole symmetric and spatially uniform. This contrasts sharply with the strongly asymmetric and spatially dependent gap-like features in Fig. 3c. This situation is again highly analogous to the homogeneous SC gap and inhomogeneous pseudogap found in underdoped cuprates[22,23].

When the doping is increased further to the non-SC regime, both the SC gap and the novel gap-like feature disappear all together. As shown in Fig. S5, the $dI/dV$ spectra in the $x = 0.109$ sample are rather featureless near $E_F$ for a wide temperature range. This behavior is characteristic of a simple metal without any ordered phase.

The large bias $dI/dV$ spectra of the NaFe$_{1-x}$Co$_x$As samples reveal another unexpected trend. As shown in Fig. 4a, the most pronounced feature between -0.5 eV and 0.5 eV is an asymmetric "V"-shaped DOS suppression around $E_F$ with a small residual DOS. A very subtle, but highly puzzling fact is that for the four samples from $x = 0$ to 0.075, the spectra remain nearly the same. The peak at negative bias always stays at -0.2 eV and the bottom of the "V"-shaped suppression is pinned at $E_F$. This behavior is certainly incompatible with the simple band structure picture, in which a rigid band shift should be induced by electron doping. Interestingly, when the doping is further increased to $x = 0.109$, the high energy spectrum suddenly starts to shift to the left (lower energy) substantially. We emphasize that the pinning of the spectra minimum at $E_F$ for $x = 0$ to 0.075 is unrelated to the SDW or SC orders at low $T$ (Fig. S6).

The complementary high energy and low energy spectra put strong constraints on the theoretical models for the iron pnictides. On one hand, the rich variations of the low energy electronic states manifest the significant role played by the itinerant electrons. On the other hand, the doping-independent high energy electronic structure indicates that the pure itinerant picture is insufficient. In fact, it has been proposed that part of the five Fe $d$ bands may be delocalized and contribute to the itinerancy, whereas the others are localized due to strong correlation effect and provide the source for local magnetic moments[7,24-27]. The schematic electronic structure for such a local-itinerant coexistent model is shown in Fig. 4b following the example in Ref. 7, in which the large "V"-shaped DOS suppression in Fig. 4a can be ascribed to a Mott-like gap for the local moment and the residual DOS can be ascribed to the itinerant bands. Initial doping into the latter should leave the large Mott-like gap unaffected and the gap bottom pinned at the $E_F$, unless the upper Hubbard band starts to get filled at strong overdoping. The magnetic susceptibility measurements of $NaFe_{1-x}Co_xAs$ shown in Fig. 4c provide another evidence for the existence of local moments. As discussed before, the linear behavior up to high $T$ cannot be explained by the itinerant electrons alone, but is consistent with the two dimensional Heisenberg model for local spins with AF interaction[28].

With the coexistence of local moments and itinerant electrons established, now we try to give a consistent picture for the complex evolution of the low energy electronic states. We start from the parent SDW state. The strongly asymmetric gap structure is energetically unfavorable for the pure FS nesting mechanism. Moreover, neutron scattering finds that the SDW wave-vector is always locked to $Q = (\pi, \pi)$ in the folded Brillouin zone notation, whereas in the FS nesting picture the SDW order is usually incommensurate and sensitive to the detailed FS geometry. However, both anomalies are unsurprising from the local-itinerant coupling point of view. The $(\pi, \pi)$ wave-vector corresponds to the AF ordering of the local moments as predicted by theory[29]. The large Hund's rule coupling between the local moments and itinerant electrons can stabilize this commensurate SDW order and lead to the large SDW gap to temperature ratio $2\Delta/k_B T_{SDW} = 9.5$, which is far beyond the conventional mean-field value[7,8]. The deviation of gap bottom from $E_F$ and the particle-hole asymmetry are due to the asymmetric Γ and M pockets with respect to $E_F$. When they are forced to shift along the (π, π) direction by electron-hole scattering, they may intersect at an energy away from $E_F$ and open a gap there, as illustrated in Fig. 4d. Thus all the unconventional features of the SDW gap can be explained naturally by the local-itinerant coupling picture.

In the optimally doped regime, the SDW gap disappears, indicating the absence of long range AF ordering. However, short range AF correlations between the local moments do exist, as seen from the linear $T$ dependence of magnetic susceptibility. It has been proposed that the short-range AF correlated local moments can support gapped spin waves, or massive magnons. Due to the Hund's rule coupling, the magnons can be emitted or absorbed by itinerant electrons via spin exchange[7,8]. The magnon-mediated electron-electron scattering provides an effective attraction in a similar manner to the BCS mechanism.

The emergence of the novel gap-like phase in the overdoped regime is totally unexpected, and there is no existing theory about it. The strong spatial variations revealed by Fig. 3c indicate that it is a local phenomenon without the involvement of long-range ordering. The localized character also rules out the simple band structure effect as a possible origin. Inspired by the similarities between this novel gap-like feature and the parent SDW gap in terms of gap size, shape, and $T$ evolution, we propose that it may originate from a similar mechanism. The main difference to the SDW state is the lack of long-range AF ordering in the overdoped regime. In fact it has been shown theoretically that the scattering between the local moments and the itinerant electrons via the Hund's rule coupling may generate a "pseudogap" with characteristic peak-dip structure at $E_F$, which is analogous to the novel gap-like feature we discovered here[30]. However, we cannot rule out the possibility that the novel gap-like feature is due to a new many body effect that has never been discussed so far.

The direct observation of three distinct gap features sheds important new lights on the nature and relationship of various phases in the iron pnictides. The fact that the optimally doped sample has neither SDW gap nor the novel gap-like feature suggests that these two phases compete with superconductivity. They may either compete for the itinerant FS sections, or the coupling modes with the local moments, or different scattering channels (electron-hole vs. electron-electron). On the other hand, the simultaneous disappearance of the SC and novel gap-like features in the strongly overdoped non-SC regime suggests that their origins are closely related. As discussed above, they both originate from the coupling between local moments and itinerant electrons, which may diminish in the strongly overdoped regime when the sinking of the hole pocket at the $\Gamma$ point invalidates the inter pocket scattering process.

## Method Summary

**Sample growth.**

High quality $NaFe_{1-x}Co_xAs$ single crystals are grown by the self flux technique. NaAs precursor is firstly synthesized at 200 °C for 10 h, and then powders of NaAs, Fe and Co are mixed together according to the ratio NaAs : Fe : Co = 4 : 1-$x$ : $x$. The mixture is placed in an alumina crucible and then sealed into an iron crucible. The samples are put in a tube furnace with inert atmosphere and melt at 950 °C for 10 h before slowly cooled down to 600 °C at a rate of 3 °C /h. The obtained crystals with shiny surface can be easily cleaved from the melt. Typical size of the as-grown crystals is 5 × 5 × 0.2 mm$^3$. The actual chemical composition of the single crystals is determined by X-ray energy dispersive spectroscopy (EDS).

**STM measurements.**

The STM experiments are performed with a low temperature ultrahigh vacuum (UHV) STM system. The NaFe$_{1-x}$Co$_x$As crystal is cleaved *in situ* at $T = 77$ K (the cleaving stage is cooled by liquid nitrogen) and then transferred immediately into the STM sample stage. An electrochemically etched polycrystalline tungsten tip is used in the experiments. The STM topography is taken in the constant current mode, and the *dI/dV* spectra are collected using a standard lock-in technique with modulation frequency $f = 423$ Hz.

**Supplementary Information** is linked to the online version of the paper.

**Acknowledgments** This work was supported by the National Natural Science Foundation of China and the Ministry of Science and Technology of China (grant No. 2009CB929400, No. 2010CB923003, and No. 2011CBA00101).

**Author Information** Correspondence and requests for materials should be addressed to Y.W. (yayuwang@tsinghua.edu.cn).

Figure Captions:

**Figure 1 | Structure and phase diagram of NaFe$_{1-x}$Co$_x$As. a,** Schematic crystal structure of NaFe$_{1-x}$Co$_x$As. It has a mirror-symmetric layered structure with two Na layers weakly bonded by the van der Waals force. The crystal is expected to cleave between these two Na layers as indicated by the grey plane. **b,** Top view of the exposed surface. **c,** The 200 Å × 200 Å constant current image of the cleaved surface of parent NaFeAs taken with sample bias $V$ = -100 mV and tunneling current $I$ = 50 pA. It displays a flat surface with several dark pits. (Inset) Close-up image taken with $V$ = 20 mV and $I$ = 50 pA shows a square lattice with 4 Å lattice constant. **d,** Schematic phase diagram of NaFe$_{1-x}$Co$_x$As. The red (blue) circle (square) marks the SC (SDW) transition temperatures of the five NaFe$_{1-x}$Co$_x$As studied in this work. **e,** The in-plane resistivity of the five NaFe$_{1-x}$Co$_x$As.

**Figure 2 | The $dI/dV$ spectra of NaFe$_{1-x}$Co$_x$As. Upper panels (a, c, e, g):** The $dI/dV$ spectra taken at $T$ = 5 K. The spectrum evolves from an asymmetric SDW gap in parent NaFeAs to a symmetric SC gap in optimally doped $x$ = 0.028 sample. The overdoped $x$ = 0.061 and 0.075 samples show the coexistence of SC gap and a large novel gap-like feature. **Lower panels (b, d, f, h):** Temperature evolution of the $dI/dV$ spectra. The blue (red) curve marks the corresponding SDW (SC) transition. The parent NaFeAs and optimally doped $x$ = 0.028 sample show a single SDW or SC gap that closes at the respective transition temperature. But the overdoped $x$ = 0.061 and

0.075 samples show a large novel gap-like feature that coexists with SC gap below $T_C$ and persists to well above $T_C$. A vertical offset is used for clarity.

**Figure 3 | Spatial distributions of the SC gap and novel gap-like feature in NaFe$_{0.939}$Co$_{0.061}$As. a,** The topography of NaFe$_{0.939}$Co$_{0.061}$As taken with sample bias $V$ = -100 mV and tunneling current $I$ = 100 pA. The red arrow marks a line on which the $dI/dV$ spectra are collected at different spots. **b,** The spectra taken at 5K along the line display spatial variation. Blue and red curves represent two typical spectra taken at the locations marked by the blue and red dots in Fig 3a. **c,** The spectra taken at 16K (above $T_C$ = 13 K) at the same locations as Fig 3b. The SC gap disappears, while the large novel gap-like feature persists and shows strong spatial variations. **d,** The normalized $dI/dV$ spectra reveal a symmetric and spatially uniform SC gap.

**Figure 4 | High energy $dI/dV$ spectra and magnetic susceptibility. a,** The high energy $dI/dV$ spectra of NaFe$_{1-x}$Co$_x$As taken at $T$ = 5 K. The DOS shows a large suppression near $E_F$, forming a large asymmetric "V" shape. The bottom of the DOS suppression is pinned at $E_F$ up to $x$ = 0.075, but shifts to the left significantly at $x$ = 0.109. **b,** Schematic electronic structure of the local-itinerant coexistent model. **c,** The $T$ dependent magnetic susceptibility of NaFe$_{1-x}$Co$_x$As. The linear $T$ dependence up to 500 K indicates the existence of local moments with AF correlation. **d,** A cartoon shows the band crossing for electron-hole scattering with ($\pi$, $\pi$) wave-vector in the folded Brillouin-zone notation. The asymmetry between the M and $\Gamma$ pocket with respect to $E_F$ gives an asymmetric SDW gap with gap bottom deviating from $E_F$.

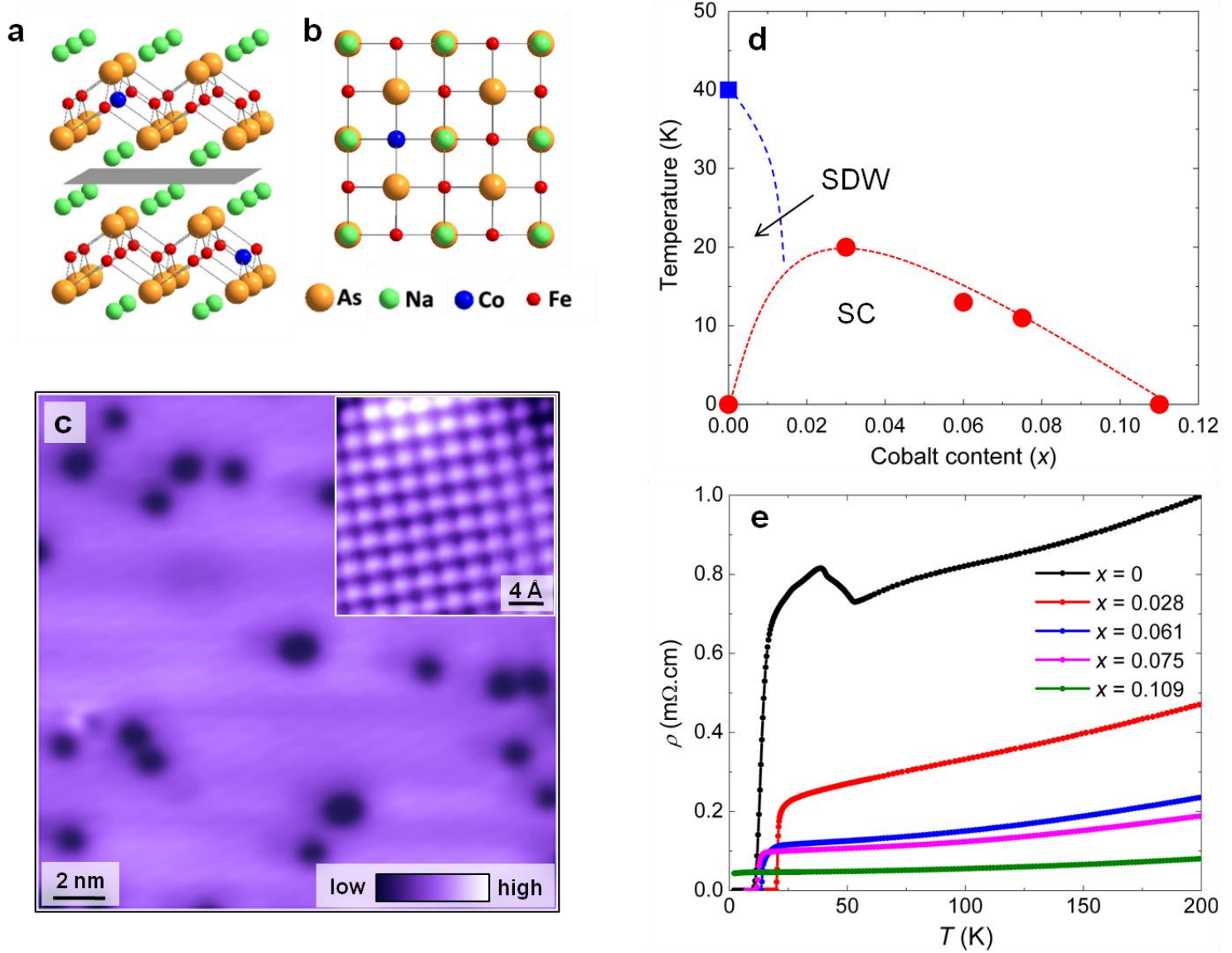

**Figure 1**

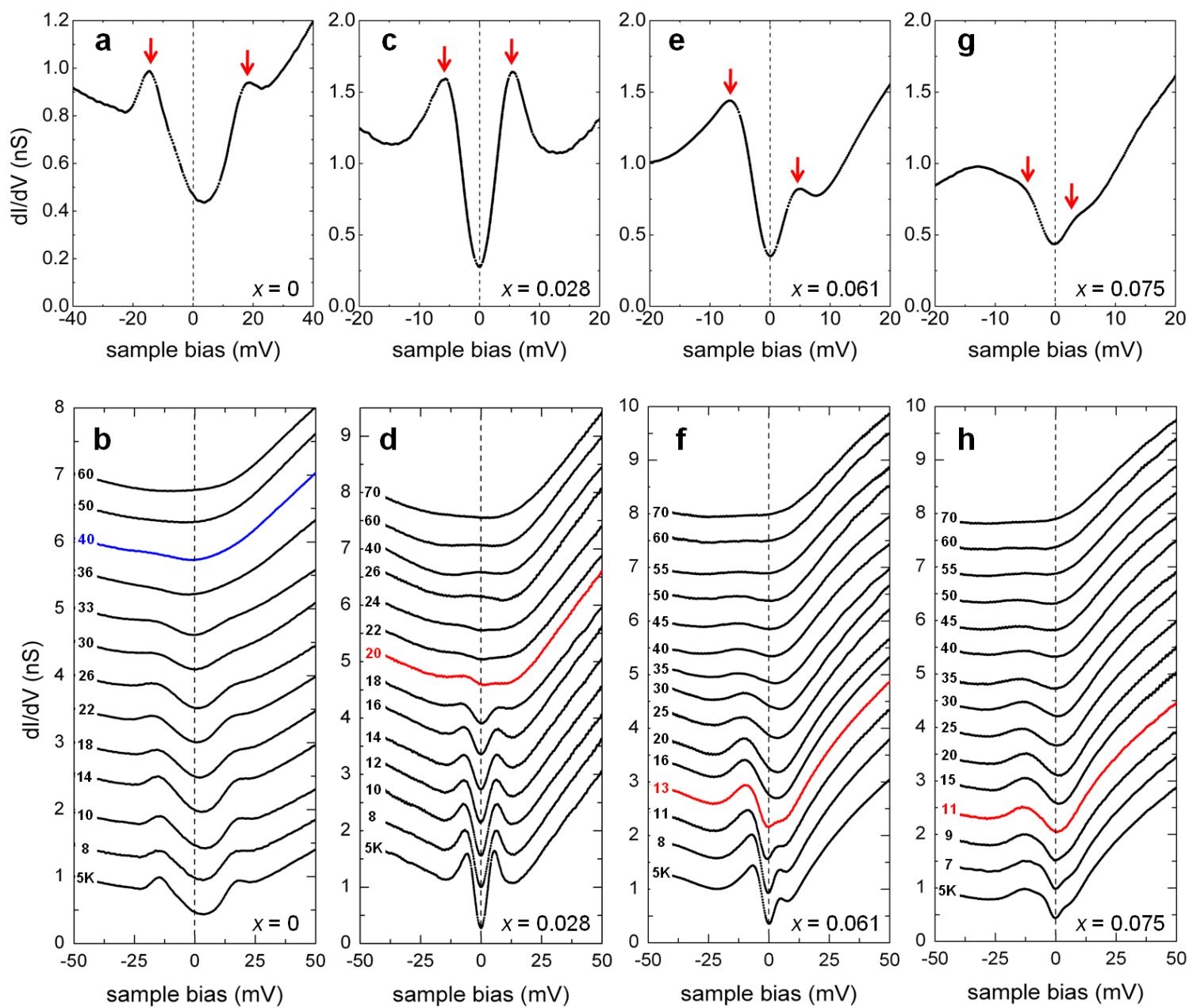

**Figure 2**

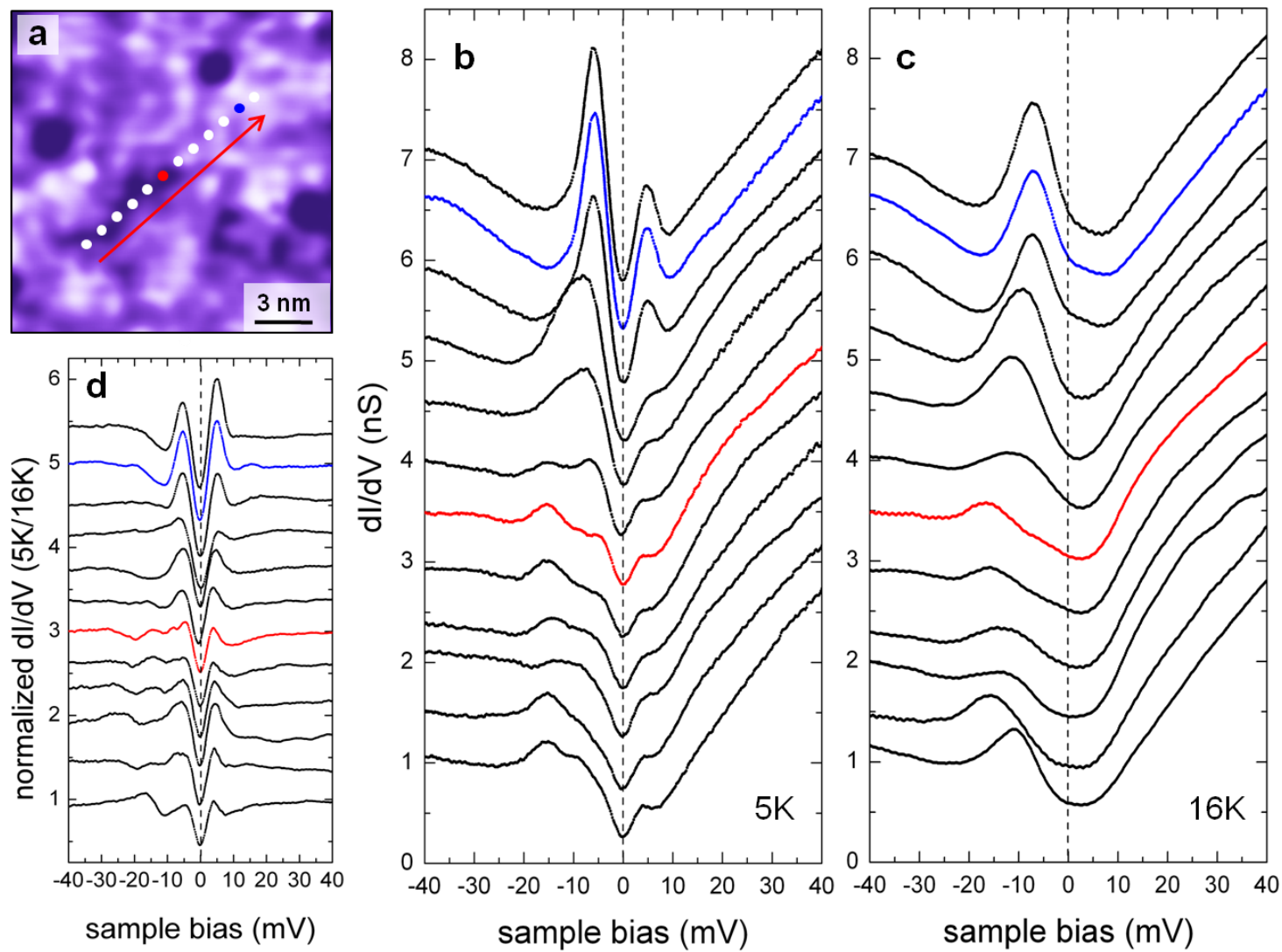

**Figure 3**

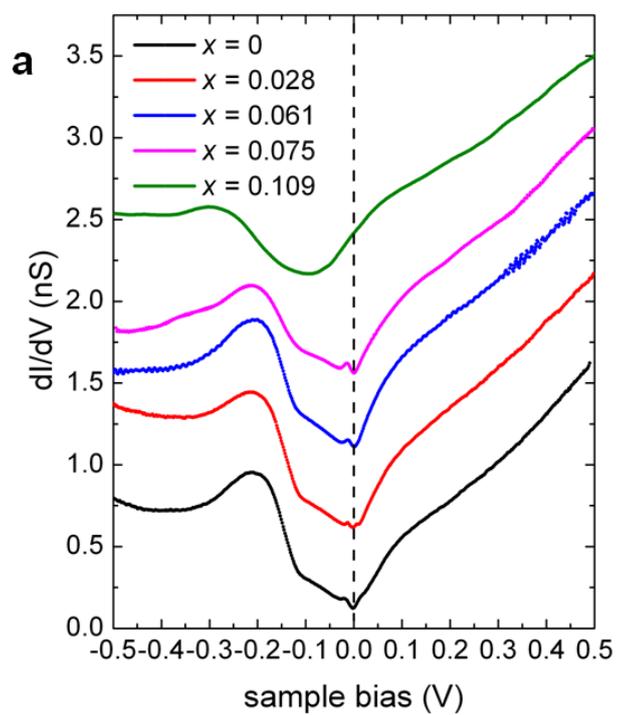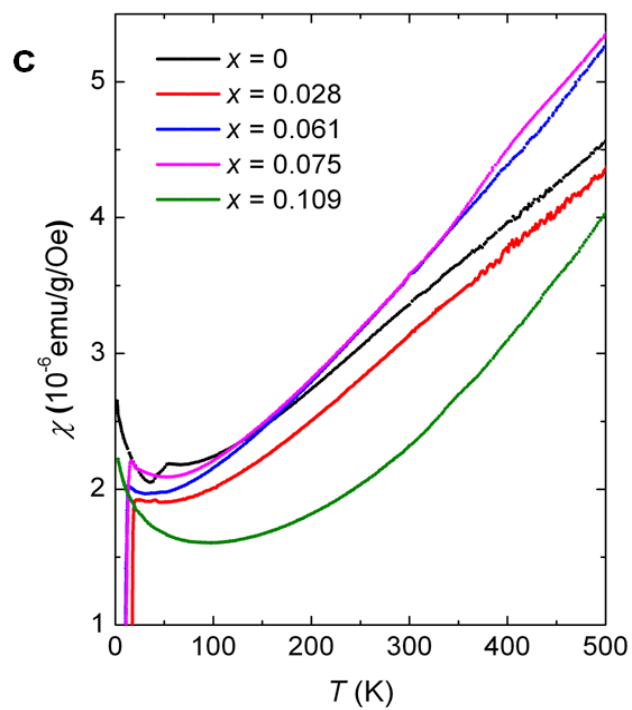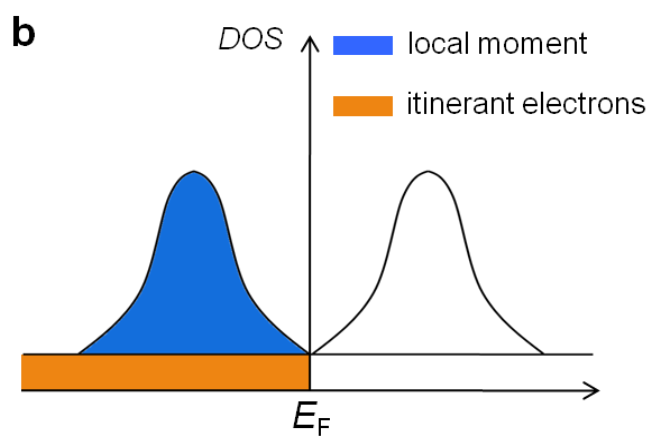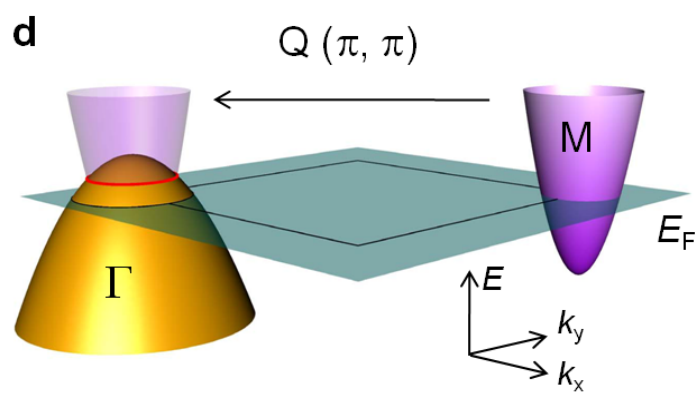

**Figure 4**

Supplementary information for

Evolution from unconventional spin density wave to superconductivity and a novel gap-like phase in NaFe$_{1-x}$Co$_x$As

Xiaodong Zhou[1,*], Peng Cai[1,*], Aifeng Wang[2], Wei Ruan[1], Cun Ye[1], Xianhui Chen[2], Yizhuang You[3], Zheng-Yu Weng[3], and Yayu Wang[1,†]

[1]*State Key Laboratory of Low Dimensional Quantum Physics, Department of Physics, Tsinghua University, Beijing 100084, P. R. China*

[2]*Hefei National Laboratory for Physical Science at Microscale and Department of Physics, University of Science and Technology of China, Hefei, Anhui 230026, P.R. China*

[3]*Institute for Advanced Study, Tsinghua University, Beijing 100084, P. R. China*

**Contents:**

**SIA: Origin of the dark pit in topography**

**SIB: Superconducting gaps obtained by the normalized *dI/dV* spectra**

**SIC: Temperature dependent normalized *dI/dV* spectra of the novel gap-like phase**

**SID: Bias voltage dependence of the topography**

**SIE: Low energy *dI/dV* spectra of the non-superconducting *x* = 0.109 sample**

**SIF: Temperature dependent high energy *dI/dV* spectra**

**Figure S1 to S6**

**References**

**SIA: Origin of the dark pit in topography**

The dark pit shown in Fig. 1c in the main text is a common topographic feature of the 111 iron pnictides surfaces[1-3]. To clarify its origin, below we present more topographic data taken on the same area of parent NaFeAs with different tunneling parameters.

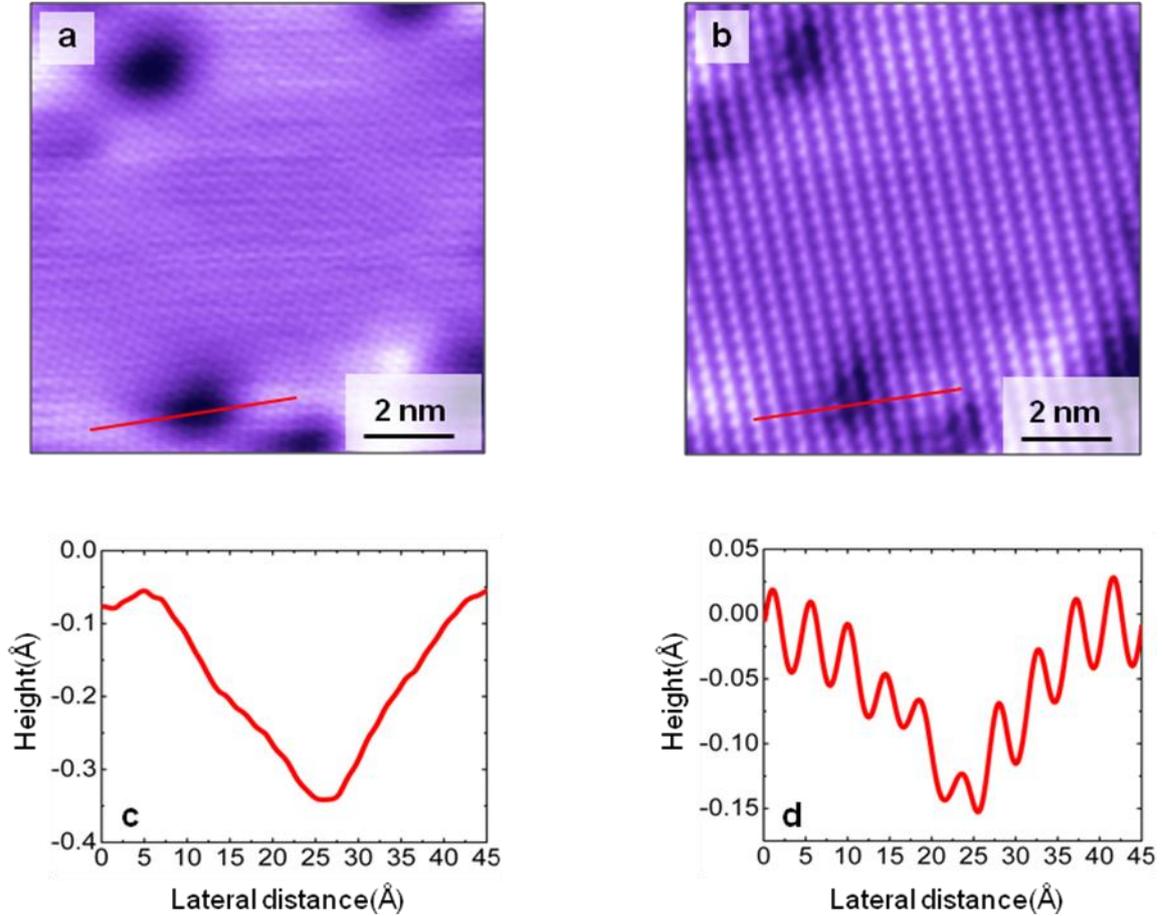

Fig. S1. **a,** The topography of parent NaFeAs taken with $V = 100$ mV and $I = 50$ pA. **b,** The topography with atomic resolution taken at the same area with $V = 15$ mV and $I = 50$ pA. **c** and **d,** Line profiles along the red lines drawn in Fig. S1a and Fig. S1b. The atomic corrugation can be seen in the dark pit (Fig. S1d), so it is not due to surface defect.

Figure S1a (taken with $V = 100$ mV and $I = 50$ pA) shows a flat surface with dark pits but without atomic resolution. In contrast, Fig. S1b (taken with $V = 15$ mV and $I = 50$ pA) shows clear atomic lattice, probably due to the closer distance between the tip and the surface. Fig. S1c and S1d are the line-cut height profiles along the red lines drawn in Fig. S1a and S1b. The atomic corrugation can be seen clearly in the dark pit area in Fig. S1d, demonstrating that there is no missing Na atom in the dark pit. So its features are most likely related to the underneath FeAs layer. The apparent "depth" of the dark pit varies from 25 pm in Fig. S1a to 10 pm in Fig. S1b, suggesting that it is not simply a structural defect but may reflect the defect electronic states of the underneath FeAs layer. However, the exact origin of the dark pit feature is still unknown at the current stage.

**SIB: Superconducting gaps obtained by the normalized *dI/dV* spectra**

To extract the superconducting (SC) gap structure, we divide the *dI/dV* spectrum measured at 5K by that taken right above the SC transition temperature $T_C$ to remove the effect of the normal state background. This is particularly necessary in the overdoped regime when the novel asymmetric gap-like feature coexists with the SC gap.

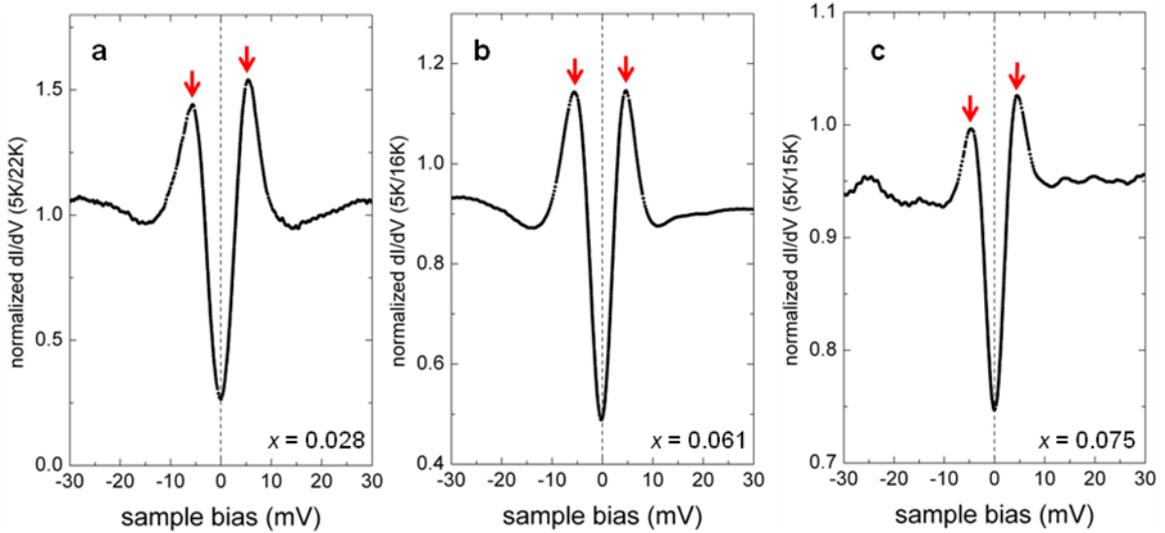

Fig. S2. **a-c,** The normalized 5K *dI/dV* spectra of NaFe$_{1-x}$Co$_x$As ($x$ = 0.028, 0.061 and 0.075).

Fig. S2a-c show the normalized 5K *dI/dV* spectra of NaFe$_{1-x}$Co$_x$As ($x$ = 0.028, 0.061 and 0.075) obtained by dividing that taken at $T$ = 22 K, 16 K, and 15 K in the three samples respectively. All the samples display a single, particle-hole symmetric SC gap with strong coherence peaks. If we estimate the SC gap value 2Δ from the spacing between the two coherence peaks, we will get Δ = 5.5 meV for $x$ = 0.028, Δ =

4.7 meV for $x = 0.061$ and $\Delta = 4.1$ meV for $x = 0.075$. The SC gap size decreases systematically with increasing doping in the overdoped regime.

**SIC: Temperature dependent normalized $dI/dV$ spectra of the novel gap-like phase**

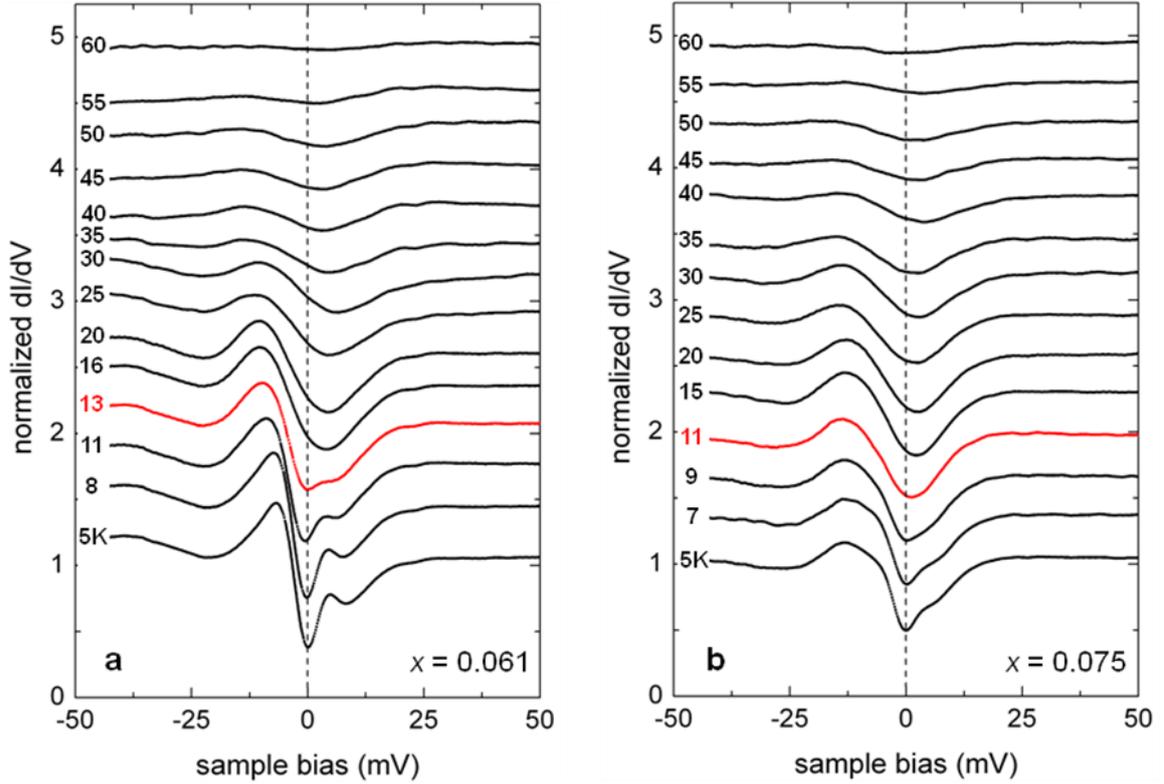

Fig. S3. The normalized $dI/dV$ spectra of overdoped NaFe$_{1-x}$Co$_x$As ($x = 0.061$ and 0.075) divided by that taken at $T = 70$ K. The red curves mark the SC transition.

To enhance the novel gap-like feature in the overdoped $x = 0.061$ and 0.075 samples, we divide the temperature dependent $dI/dV$ spectra by the one taken at 70 K to remove the effect of the high $T$ background. The normalized spectra shown in Fig. S3a and S3b clearly reveal that the novel gap-like feature coexists with the SC gap below $T_C$ and persists deep into the normal state. It gradually closes at around $T = 55$ K. The overall behavior is highly analogous to the pseudogap in the cuprates[4].

**SID: Bias voltage dependence of the topography**

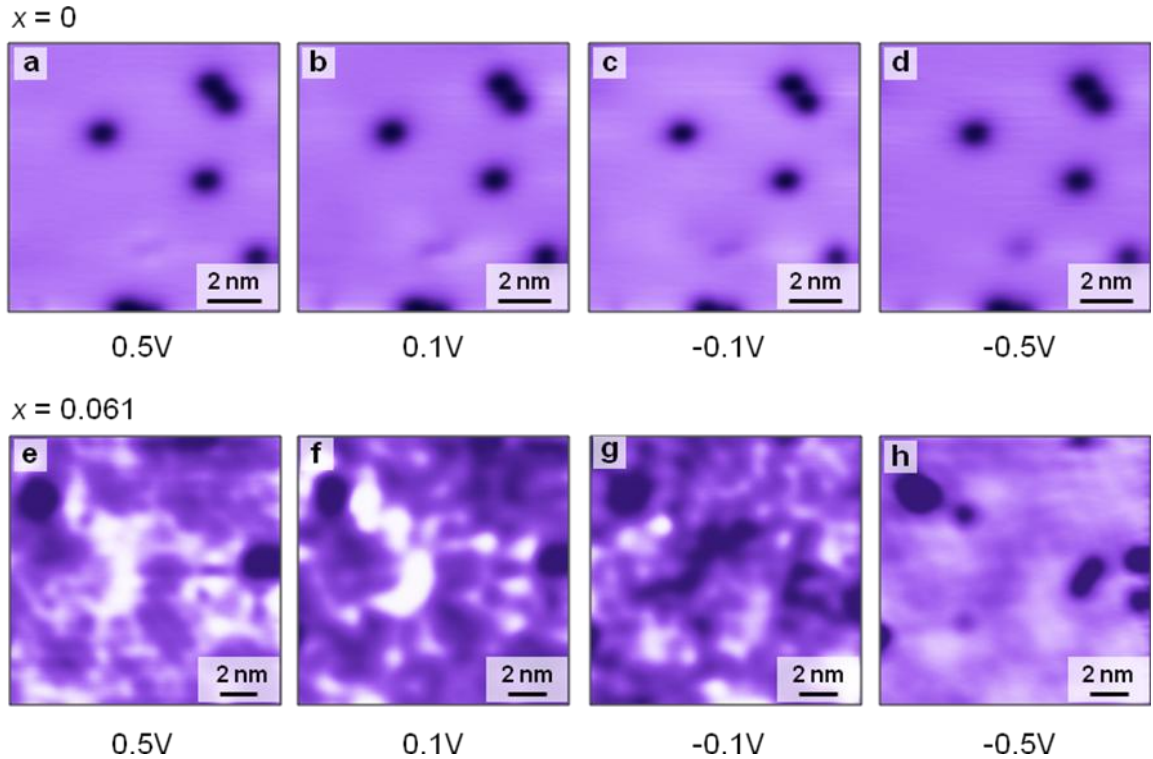

Fig. S4. **a, b, c, d,** The topographic images of parent NaFeAs taken at 4 different bias voltages. The surface is clean and shows weak bias voltage dependence. **e, f, g, h,** The topographic images of the overdoped $x = 0.061$ sample taken at 4 bias voltages. It becomes more disordered due to Co doping and shows strong bias voltage dependence.

Figure S4 display the topographic images of the parent NaFeAs and overdoped $x = 0.061$ sample taken at 4 different bias voltages. The topography of parent NaFeAs shows clean surface with dark pits, and the features have very weak bias-voltage dependence (see Fig. S4a-d). The overdoped $x = 0.061$ samples, on the other hand, are much more disordered (see Fig. S4e-h). The disorder features are strongly bias dependent, suggesting that they are due to the impurity electronic states induced by the Co substation of Fe in the underneath FeAs plane. At -0.5V bias voltage, the surface is rather smooth, similar to that in parent NaFeAs without Co doping, suggesting that the surface structure is little affected by the Co doping.

**SIE: Low energy *dI/dV* spectra of the non-superconducting $x = 0.109$ sample**

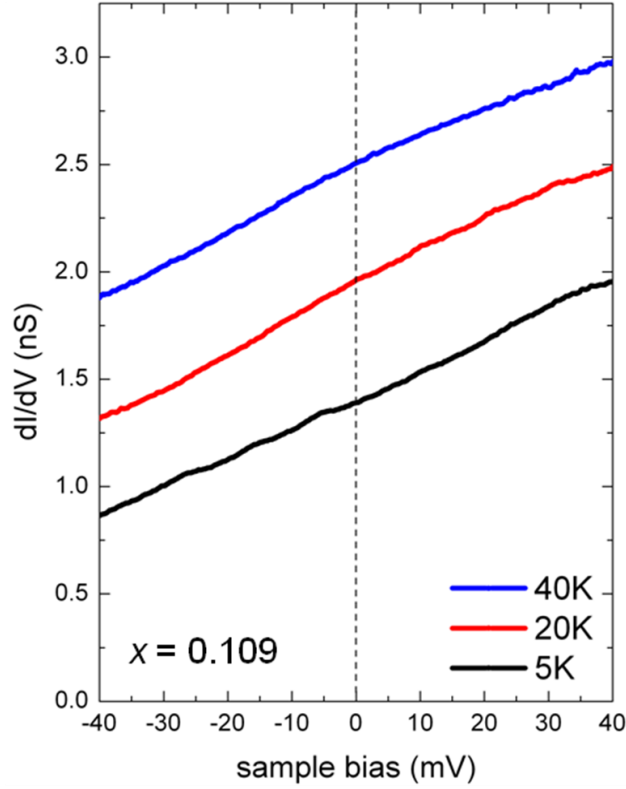

Fig. S5. The low energy *dI/dV* spectra of the strongly overedoped, non-SC $x = 0.109$ sample show no sign of the SC gap and the novel gap-like feature.

Figure S5 shows the temperature dependent low energy *dI/dV* spectra of the strongly overdoped non-SC $x = 0.109$ sample. The spectra are featureless near the Fermi energy with no temperature dependence, which are typical for normal metals without any ordering. Both the SC gap and the novel gap-like feature are absent in this sample.

**SIF: Temperature dependent high energy *dI/dV* spectra**

In Fig. 4a in the main text, we display the large bias (high energy) spectra of the five NaFe$_{1-x}$Co$_x$As samples measured at $T = 5$ K. The most striking feature there is the four samples with $x = 0$ to $0.075$ have nearly the sample high energy electronic structure. Especially, the bottom of the large DOS suppression is pinned at the Fermi level in this doping range, apparently violating the simple band structure picture.

However, since these spectra are taken at low $T$ when the samples are in the SDW or SC states, it is unclear if the pinning effect is due to the formation of the low energy ordered phase. In order to clarify this point, we extend the high energy spectroscopy measurements to temperatures higher than the respective phase transition temperature.

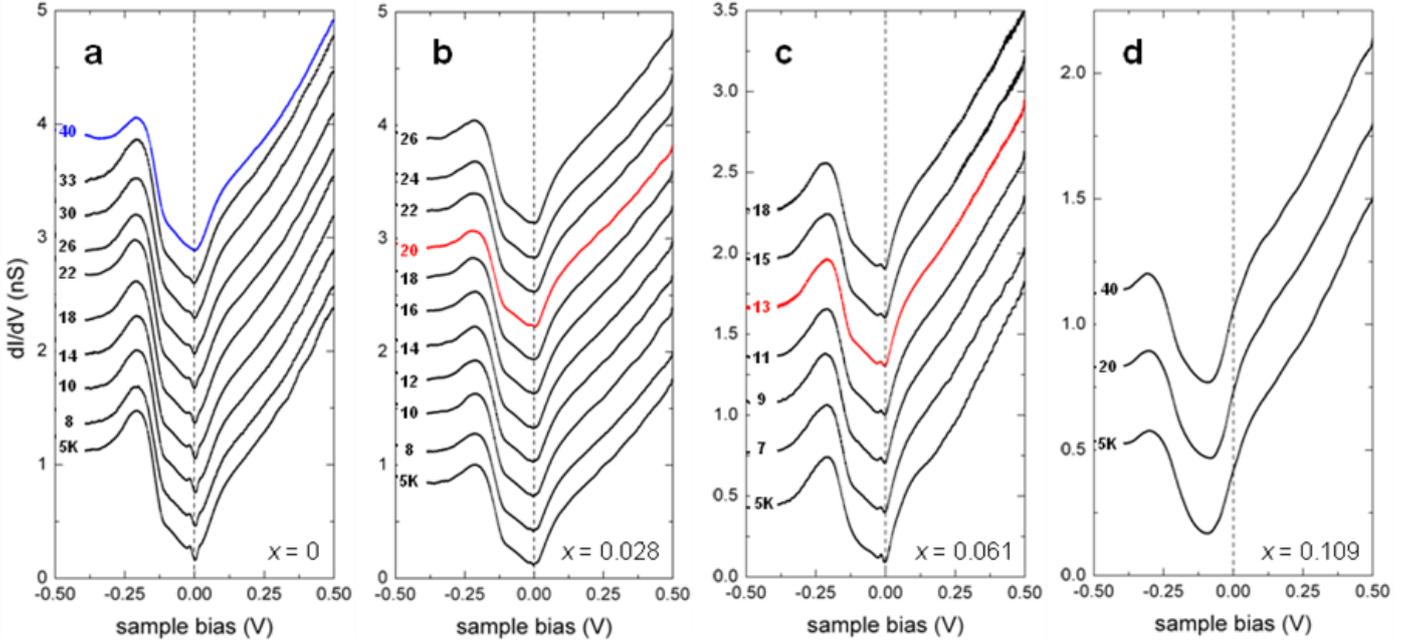

Fig. S6. The high energy $dI/dV$ spectra of NaFe$_{1-x}$Co$_x$As ($x$ = 0, 0.028, 0.061 and 0.109) measured at varied temperatures. The blue and red curves mark the SDW or SC transitions. The spectra are offset vertically for clarity.

Shown in Fig. S6a-d are the high energy $dI/dV$ spectra of four NaFe$_{1-x}$Co$_x$As ($x$ = 0, 0.028, 0.061 and 0.109) measured at varied temperatures. As can be seen clearly in Fig. S6a to S6c, for the three samples from $x$ = 0 to 0.061 the minimum of the large "V"-shaped DOS suppression is still pinned at $E_F$ even at temperatures above their respective transition temperatures. It is not unexpected because the large energy scale electronic structure should not be affected by the change over such a small temperature range. This "pinning" effect is only lifted in the strongly overdoped non-SC $x$ = 0.109 sample (Fig. S6d), in which the "V"-shaped DOS suppression is still present but its minimum now moves to -95 meV below $E_F$. The overall features cannot be explained by a simple band structure picture, and strong correlation effect has to be considered.